\documentclass[12pt]{article}
\usepackage{graphicx}
\begin{document}
\noindent {\footnotesize\it  
 Astronomy Letters, 2009, Vol. 35, No. 6, pp. 396--405. 
 }

\noindent
\begin{tabular}{llllllllllllllllllllllllllllllllllllllllllllll}
 & & & & & & & & & & & & & & & & & & & & & & & & & & & & & & & & & & & & &  \\\hline\hline
\end{tabular}

 \vskip 1.5cm
 \centerline {\large\bf Open Clusters ASCC21 as a Probable Birthplace}
 \centerline  {\large\bf of the Neutron Star Geminga}
 \bigskip
 \centerline {V.V. Bobylev and A.T. Bajkova}
 \bigskip
 \centerline {\small\it
 Pulkovo Astronomical Observatory, Russian Academy of Sciences, St-Petersburg}
 \bigskip

{\bf Abstract--}We analyze the encounters of the neutron star
(pulsar) Geminga with open star clusters in the OB association
OriOB1a through the integration of epicyclic orbits into the past
by taking into account the errors in the data. The open cluster
ASCC21 is shown to be the most probable birthplace of either a
single progenitor star for the Geminga pulsar or a binary
progenitor system that subsequently broke up. Monte Carlo
simulations of Geminga--ASCC21 encounters with the pulsar radial
velocity $V_r = -100\pm50$ km s$^{-1}$ have shown that close
encounters could occur between them within $\leq 10$ pc at about
$t=-0.52$ Myr. In addition, the trajectory of the neutron star
Geminga passes at a distance of $\approx$25 pc from the center of
the compact OB association $\lambda$~Ori at about $t=-0.39$ Myr,
which is close to the age of the pulsar estimated from its timing.

\section*{INTRODUCTION}

Young massive O and B stars can run away from the stellar group
(an open cluster or an OB association) where they were born. Such
stars are called runaways; they are identified by high ($>30$ km
s$^{-1}$) space velocities. Two main runaway scenarios are known
(see the review by Hoogerwerf et al. 2001). The first scenario is
related to the evolution of a binary system and the explosion of
one of the components as a supernova. In this case, one of the
components can receive a high kick velocity due to an asymmetric
explosion and a significant mass loss by the binary. A massive
runaway star and its former companion, a neutron star or a black
hole, are the breakup products of such a binary. The second
scenario is related to the dynamical evolution of an initial
stellar group. Mutual encounters of close binary and multiple
systems are the most efficient ejection mechanism.

The best-known runaway stars, such as $\zeta$Oph, $\xi$Per,
$\mu$Col, AE Aur, $\zeta$Pup, and others, were discovered by
Blaauw~(1961; 1993). Probable parent associations were determined
for some of them. Hoogerwerf et al. (2001) considered the problem
of determining the most probable birthplaces of pulsars using data
from the Hipparcos (1997) Catalogue. Based on a  more complete
list of runaway OB stars, Schilbach and R\"oser (2008) showed that
the parent open cluster or OB association to which they belonged
could be pinpointed with a high probability for most
($\approx90\%$) of the O stars in the solar neighborhood (within
$<1$ kpc), including those that were previously believed to be
field stars.

The problem of determining probable birthplaces of nearby young
neutron stars associated with the Gould Belt (Popov et al. 2003;
Motch et al. 2006) is of great interest. One of these stars is
Geminga (PSR J0633$+$1746), to which this paper is devoted.

In contrast to the overwhelming majority of pulsars, Geminga was
discovered in 1973 as a gamma-ray source from the SAS-2 satellite.
The impulsive behavior of its signals was first detected in the
X-ray (Halpern and Holt 1992) and gamma-ray (Burch et al. 1992;
Bignami and Caraveo 1992) bands and, only several years later, in
the radio band (Kuzmin and Losovsky 1997; Malofeev and Malov
1997). The age of the Geminga pulsar, 0.342 Myr, was first
determined from its timing (P/2\.P , where P and \.P are the
pulsar period and its first derivative, respectively) by Bignami
and Caraveo (1992). Bignami et al. (1993) identified this pulsar
in the optical band with a 25.5-magnitude object and were the
first to determine the components of its proper motion. The
distance to the Geminga pulsar, $r=150-400$ pc, was first
estimated by Halpern and Ruderman (1993).

The present-day quality of the spectra for pulsars is too low for
their radial velocities to be determined. At the same time, an
arc-like feature produced by the interaction of pulsar emission
with the interstellar medium is observed near the neutron star
Geminga (Caraveo et al. 2003; Shibanov et al. 2006). Modeling the
shape of this arc-like feature (Caraveo et al. 2003) yielded a
constraint on the magnitude of the radial velocity for the Geminga
pulsar ($|V_r|\leq100$ km s$^{-1}$). Given this constraint, the
radial velocity is a free parameter in our studies.

One of the first hypotheses that the birthplace of the Geminga
pulsar is associated with the Local Bubble was proposed by Gehrels
and Chen (1993). According to this hypothesis, a positive radial
velocity of the pulsar should be adopted. In the opinion of Frisch
(1993), this pulsar originated in the Orion OB association,
suggesting that its radial velocity is negative. The modeling by
Smith et al. (1994) showed that the compact OB association
$\lambda$Ori (Cr 69) at $V_r\approx-100$ km s$^{-1}$ could be a
suitable birthplace of the pulsar.

Using Hubble Space Telescope (HST) observations, Caraveo et al.
(1996) were the first to determine the trigonometric parallax of
the pulsar, $\pi_{tr} = 6.36\pm1.74$ mas, which corresponds to its
distance $r = 157^{+59}_{-34}$ pc. They also determined the proper
motion components, $\mu_\alpha\cos\delta=138\pm4$ mas yr$^{-1}$
and $\mu_\delta = 97\pm4$ mas yr$^{-1}$. The search for a probable
birthplace of the neutron star Geminga carried out by Pellizza et
al. (2005) based on these data led them to conclude that this
birthplace could be in a fairly close solar neighborhood ($r =
90-240$ pc) and is related either to the relatively old
($\approx$50 Myr) Cas傍au association or to the edge of the
younger ($\approx$11 Myr) OB association Ori OB1a.

The currently available HST observations of Geminga (Faherty et
al. 2007) showed that the new parallax, $\pi_{tr} = 4.0\pm1.3$ mas
($r = 250^{+120}_{-62}$ pc), differs significantly from the
previously published one (Caraveo et al. 1996), which should
inevitably affect the results of searching for its birthplace.

The goal of this paper is to determine the most probable
birthplace of the Geminga pulsar by simulating its motion in space
into the past using the most resent data on open star clusters
(Kharchenko et al. 2007) and individual stars (van Leeuwen 2007).
The objectives set here are accomplished using the method of
statistical Monte Carlo simulations--we determine the epicyclic
orbits of objects by taking into account the corridor of errors in
the observational data.

{
\begin{table}[t]
\caption[]{\small
 Data on stars and open star clusters Object.

}
\begin{center}
\begin{tabular}{|l|c|c|c|c|c|c|c|c|c|c|}\hline
         &&&&&\\
 Object  & $\alpha_{(J2000.0)},$ & $\mu_\alpha\cos \delta,$~~~~~ & $\pi,$ & $V_r,$ & Distance,   \\
         & $\delta_{(J2000.0)}$  & $\mu_\delta,$  mas yr$^{-1}$  &   mas  & km s$^{-1}$  & pc      \\\hline

 ASCC16 & $5^h 24^m 36^s$    & $ 0.75\pm0.22$ &  & $20.80\pm4.60$ & $460$   \\
        & $1^\circ 48' 00''$ & $-0.18\pm0.29$ &  &                & (2)     \\\hline

 ASCC18 & $5^h 26^m 09.6^s$  & $ 0.89\pm0.28$ &  & $ 24\pm10$     & $500$   \\
        & $0^\circ 49' 12''$ & $-0.02\pm0.25$ &  &                & (2)     \\\hline

 ASCC20 & $5^h 28^m 44.4^s$  & $-0.09\pm0.21$ &  & $22.97\pm4.80$ & $450$   \\
        & $1^\circ 37' 48''$ & $ 0.51\pm0.19$ &  &                & (2)     \\\hline

 ASCC21 & $5^h 28^m 58.8^s$  & $ 0.52\pm0.25$ &  & $19.77\pm1.12$ & $500$   \\
        & $3^\circ 39' 00''$ & $-0.62\pm0.28$ &  &                & (2)     \\\hline

 $\lambda$~Ori & $ 5^h 35^m 09.6^s$  & $ 0.40\pm0.47$ &  & $31.38\pm1.42$ & $438$   \\
 (Cr~69)       & $9^\circ 42' 00''$  & $-1.93\pm0.34$ &  &                & (2)     \\\hline

 HIP 22061    & $4^h 44^m 42^s.1571$   & $-44.02\pm0.52$ & $2.99\pm0.63$ & $9.0\pm4.4$ & $$  \\
              & $0^\circ 34' 05''.418$ & $-30.04\pm0.50$ & (1)           &  (3) &           \\\hline

 HIP 29678    & $ 6^h 15^m 08^s.4567$   & $25.23\pm0.50$ & $2.72\pm0.43$ & $35.8\pm1.0$ & $$  \\
              & $13^\circ 51' 03''.859$ & $11.44\pm0.30$ & (1)           &  (4) &           \\\hline

 Geminga & $6^h 33^m 54^s.1530$       & $142.2\pm1.2$ & $4.0\pm1.3$ &   &  \\
         & $17^\circ 46' 12''.909$    & $107.4\pm1.2$ &   (5)       &   &  \\\hline
\end{tabular}
\end{center}
 {\small

Note. (1) van Leeuwen (2008), (2) CRVOCA (Kharchenko et al. 2007),
(3) PCRV (Gontcharov 2006), (4) CRVAD-2 (Kharchenko et al. 2007),
(5) Faherty et al. (2007).
 }
\end{table}
}

\section*{DATA}

The input data on the objects under consideration, such as their
equatorial coordinates, proper motion components, radial
velocities, and parallaxes, are given in the table. The data for
all of the stars used were taken from a revised version of the
Hipparcos Catalogue (van Leeuwen 2007). For the Geminga pulsar, we
use new data from Faherty et al. (2007), which were obtained from
18-month-long HST observations of the star. The CRVOCA Catalog
(Kharchenko et al. 2007) served as the source of data for open
clusters. The distance estimates in this catalog are based on a
photometric method with an error of $\approx20\%$. We calculated
the parallaxes for each open cluster based on the distances given
in the last column of the table.

Four groups,Ori OB1a, b, c, and d (Blaauw1964), with ages of
$11.4\pm1.9, 1.7\pm1.1, 4.6\pm2,$ and $<1$~Myr, respectively, are
identified in the Orion association (Brown et al. 1994). According
to Kharchenko et al. (2005), the Ori OB1a association includes
four open clusters, ASCC16, ASCC18, ASCC20, and ASCC21, with ages
of 8.5, 13.2, 22.4, and 12.9 Myr, respectively. The question of
whether each of them is an open cluster or an association is still
under discussion. In this respect, ASCC16, which is also
designated as the group 25 Ori (Brice\~{n}o et al. 2005; McGehee
2006), is the best-studied open cluster. Here, a fairly dense
concentration of young T Tauri stars was detected. Their analysis
led McGehee (2006) to conclude that this is a gravitationally
unbound group rather than an open cluster.

The OB--T association/cluster $\lambda$Ori is located near the Ori
OB1a association. A compact HII region (S264) is associated with
the star $\lambda$Ori. In a wider neighborhood of the center with
a radius of $\approx5^\circ$ (40 pc), the association is
surrounded by a ring of molecular clouds and gas (Maddalena and
Morris 1987). This star-forming region was studied in a series of
papers by Dolan and Mathieu (1999, 2001, 2002). Based on
Str\"omgren photometry, they estimated the distance to the open
cluster $\lambda$Ori, $450\pm50$ pc. Ten B stars forming the
cluster proper were shown to concentrate in a sky field with a
radius of $\approx1^\circ$ (8 pc) around the central star of the
cluster $\lambda$Ori (Sp: O8III) with an age of $\approx$5.5 Myr.
In a wider neighborhood of the center with a radius of
$\approx5^\circ$ (40 pc), they detected several concentrations of
young T Tauri stars. On the whole, Dolan and Mathhieu concluded
that: (a) star formation in the cluster $\lambda$Ori ceased
$\approx$1 Myr ago after the supernova explosion; (b) the group of
young massive stars is currently not gravitationally bound; and
(c) star formation currently goes on in the clouds B30 and B35
located on the periphery of the association.

\section*{METHODS}
\subsection*{Orbit Construction}

In this paper, we use a rectangular Galactic coordinate system
with the axes directed away from the observer toward the Galactic
center ($l=0^\circ, b=0^\circ,$ the $X$ axis), along the Galactic
rotation ($l=90^\circ, b=0^\circ,$ the $Y$ axis), and toward the
North Galactic Pole ($b=90^\circ,$ the $Z$ axis). The
corresponding space velocity components of the objects $U,V,W$ are
also directed along the $X,Y,Z$ axes (Kulikovskii 1985).

The epicyclic approximation (Lindblad 1927, 1959) allows the
orbits of stars to be constructed in a coordinate system rotating
around the Galactic center. We assume that the origin of the
coordinate system coincides with the local standard of rest and
that the stars move along epicycles in the direction opposite to
the Galactic rotation. We use the method in the form given by
Fuchs et al. (2006):
$$
\displaylines{\hfill
 X(t)= X(0)+{U(0)\over \kappa} \sin(\kappa t)
      +{V(0)\over 2B} (1-\cos(\kappa t)),         \hfill\llap(1)\cr\hfill
 Y(t)= Y(0)+2A \biggl( X(0)+{V(0)\over 2B}\biggr) t
       -{\Omega_0\over B\kappa} V(0) \sin(\kappa t)
       +{2\Omega_0\over \kappa^2} U(0) (1-\cos(\kappa t)),\hfill\cr\hfill
 Z(t)= {W(0)\over \nu} \sin(\nu t) + Z(0) \cos(\nu t), \hfill
 }
$$
where $t$ is the time in Myr (pc/Myr=0.978 km s$^{-1}$), which is
measured into the past; $A$ and $B$ are the Oort constants;
$\kappa=\sqrt{-4\Omega_0 B}$ is the epicyclic frequency;
$\Omega_0$ is the angular velocity of Galactic rotation of the
local standard of rest, $\Omega_0= A-B$; and $\nu=\sqrt{4\pi G
\rho_0}$ is the frequency of the vertical oscillations, where $G$
is the gravitational constant and $\rho_0$ is the star density in
the solar neighborhood. The space velocities of the objects are
calculated for any necessary time using the formulas
$$ \displaylines{\hfill
 U(t)= U(0) \cos(\kappa t)+{\kappa\over 2B} V(0) \sin(\kappa t), \hfill\llap(2)\cr \hfill
 V(t)= 2A \biggl(X(0)+{V(0)\over 2B}\biggr)
       -{\Omega_0\over B} V(0) \cos(\kappa t)
       +{2\Omega_0\over \kappa} U(0) \sin(\kappa t)),\hfill\cr\hfill
 W(t)= W(0) \cos(\nu t) - Z(0)\nu \sin(\nu t). \hfill }
$$
The parameters $X(0),Y(0),Z(0)$ and $U(0),V(0),W(0)$ in Eqs. (1)
and (2) denote, respectively, the current positions and velocities
of the objects. The velocities $U,V,W$ are given relative to the
local standard of rest with $(U,V,W)_{LSR}=(10.00,5.25,7.17)$ km
s$^{-1}$ (Dehnen and Binney 1998). Following Fuchs et al. (2006),
we adopted $\rho_0 = 0.1 M_\odot$ pc$^3$, which gives $\nu=74$ km
s$^{-1}$ kpc$^{-1}$. We also used the Oort constants
$A=13.7\pm0.6$ km s$^{-1}$ kpc$^{-1}$ and $B= -12.9\pm0.4$ km
s$^{-1}$ kpc$^{-1}$ that we found previously (Bobylev 2004) by
analyzing the independent determinations of these parameters by
various authors; $\kappa = 37$ km s$^{-1}$ kpc$^{-1}$ corresponds
to these values.

\subsection*{Statistical Simulations}

In accordance with the method of statistical Monte Carlo
simulations, we compute 3 million orbits for each object by taking
into account the random errors in the input data. For each pair of
orbits belonging to two different objects, we calculate the
encounter parameter equal to the minimum separation between the
objects $\Delta r=\sqrt{\Delta X^2(t)+\Delta Y^2(t)+\Delta
Z^2(t)}$.

The parameters of the objects are distributed normally with a
dispersion $\sigma$. As was shown by Hoogerwerf et al. (2001), the
random errors in the equatorial coordinates of stars and open
clusters do not affect noticeably the simulation results. As a
result, the errors are added only to the proper-motion components,
parallax, and radial velocity of an object. The parameters are
listed in the table.

The distribution of the encounter parameter in a certain
neighborhood can be represented as a histogram (see below). We
calculated the expected distribution $F_{3D}$ of the minimum
separation $\Delta_r$ using a formula from Hoogerwerf et al.
(2001),
$$ \displaylines{\hfill F_{3D}(\Delta_r)=
 {\Delta_r\over {2\sigma\mu\sqrt\pi}}
 \biggl\{\exp\biggl[-{(\Delta_r-\mu)^2\over {4\sigma^2}}\biggr]
         -\exp\biggl[-{(\Delta_r+\mu)^2\over {4\sigma^2}}\biggr]
 \biggr\} \hfill\llap(3)}
$$
for appropriate mean $\mu$ and dispersion $\sigma$. When $\mu$ may
be considered close to zero, Hoogerwerf et al. (2001) obtained the
expression
$$ \displaylines{\hfill F_{3D}(\Delta_r)=
 {\Delta^2_r\over {2\sigma^3\sqrt\pi}}\exp\biggl
 [-{\Delta^2_r\over{4\sigma^2}}\biggr]. \hfill\llap(4)}
$$
For such open clusters as ASCC21 and $\lambda$Ori, the tidal
radius is $13.7\pm1.8$ and $7.7\pm1.4$ pc, respectively (Piskunov
et al. 2008). Therefore, in our simulations, we choose the
characteristic radius of the neighborhood in which close
encounters of stars with clusters occur to be 10 pc.

Analysis of the orientation of the shock front ahead of the pulsar
yields a constraint on its radial velocity. A fairly symmetric
shock was detected near the neutron star Geminga (Caraveo et al.
2003; Shibanov et al. 2006). Modeling its shape suggests that the
radial velocity of Geminga is relatively low, $|V_r|\leq 72$ km
s$^{-1}$; the angle $\beta$ between the total space $(V)$ velocity
vector and the tangential $(V_\bot)$ velocity is $|\beta|\leq
30^\circ$ (Caraveo et al. 2003; Pellizza et al. 2005). For the new
data (table), $\beta = 30^\circ.7,$ where $|\cos \beta| =
V_\bot/V,$ corresponds to the radial velocity $|V_r| = 100$ km
s$^{-1}$. Therefore, for the Geminga pulsar, we take its model
radial velocity $V_r = -100\pm50$ km s$^{-1}$.

\section*{RESULTS}

Figure 1 shows the positions and trajectories of the open clusters
in the Ori OB1a association, the cluster $\lambda$Ori, and the
neutron star Geminga for a radial velocity of $-100\pm50$ km
s$^{-1}$ over 1 Myr into the past (with marks at 0.5-Myr
intervals). To construct the trajectories in Galactic coordinates,
we first calculated the rectangular coordinates $X(t),Y(t),$ and
$Z(t)$ from Eqs. (1) and then found the corresponding spherical
Galactic coordinates $l(t)$ and $b(t)$ from them. As can be seen
from the figure, the trajectory of the Geminga pulsar passes near
two open clusters, ASCC21 and $\lambda$Ori. The two dotted lines
in Fig.~1 indicate the pulsar trajectories calculated with extreme
$(\pm3\sigma)$ values of the proper-motion components.

Our simulations of the Geminga encounter with the cluster ASCC21
yielded the following results. Out of the 3 million orbits
obtained by introducing random errors with a given dispersion for
both pulsar and cluster, 10 5441 encounters occur at separations
$\Delta_r\leq10$ pc $(3.5\%)$. In 3137 of the 10 5441 cases
$(3.0\%)$, the encounters at about $t = -0.52$ Myr occur in the
neighborhood of ASCC21 with a radius $\leq10$ pc and with the
center lying on the cluster trajectory obtained from the data
without introducing any errors. Figure~2a shows the expected
distribution $F_{3D}$ of the minimum separation $\Delta_r$
calculated from Eq.~(3) for the adopted mean $\mu = 7.75$ pc and
dispersion $\sigma = 1.0$ pc. The domain of admissible values of
$V_r$, $\pi$, $\mu_\alpha\cos\delta$ ($\mu^*_\alpha$), and
$\mu_\delta$ at which 3137 encounters occur are shown in Fig. 3
for the Geminga pulsar and Fig.~4 for the cluster ASCC21. The
horizontal and vertical lines in Figs.~3 and 4 pass through the
nominal values of the parameters used.

Figure~3 shows the set of lines representing the region that
satisfies the condition $|\beta|\leq30^\circ$ for the $V_r-\pi$
relation. The lines were obtained in accordance with the relation
$V_r = V\bot \tan \beta.$ The solid and dotted lines correspond to
the angles $\beta=\pm30^\circ$ and $\beta=\pm15^\circ,$
respectively. As we see from the figure, most of the model points
fall into the interval $|\beta|\leq30^\circ,$ thereby satisfying
the constraint on the pulsar radial velocity determined from an
analysis of the shock shape (Caraveo et al. 2003; Pellizza et al.
2005).

Our simulations of the encounter of the neutron star Geminga with
the cluster $\lambda$Ori revealed no mutual encounters at
separations $\leq$10 pc. However, there are less close encounters.
Thus, for example, out of the 300 000 model orbits, 25 745
encounters occur at separations $\Delta_r\leq25$ pc $(8.5\%);$ in
2686 of the 25 745 cases, the encounters occur within $\leq$30 pc
$(10\%)$ of the cluster $\lambda$Ori at about $t = -0.39$ Myr.

Our simulations of the Geminga encounter with the cluster ASCC16
yielded the following results: out of the 3 million orbits, 27 848
encounters occur at separations $\Delta_r\leq10$ pc $(0.9\%);$ in
1475 of the 27 848 cases, the encounters occur within $\leq$50 pc
of the cluster ASCC16 at about $t = -0.63$ Myr. Thus, in this
case, we have a factor of 4 fewer encounters than in the case of
ASCC21.

To determine a possible candidate for the binary component with
the neutron star Geminga in the past, we selected known rapidly
flying stars whose trajectories passed near the Ori OB1
association on the celestial sphere in a time interval of 2 Myr
(Fig. 2 from Hoogerwerf et al. 2001): HIP 22061, HIP 27204, and
HIP 29678 (numbers 4, 6, and 7 in the list of Hoogerwerf et al.,
2001).

For HIP 22061, there are encounters with the Geminga pulsar at
$t\approx0.6$ Myr but within $\Delta_r\leq20$ pc, i.e., the
encounters are not close. For HIP 27204, there are no encounters
with the Geminga pulsar within $\Delta_r\leq20$ pc. For HIP 29678,
there are close encounters with the Geminga pulsar. We found that
out of the 3 million orbits, 124 544 encounters occurred at
separations $\Delta_r\leq10$ pc $(4\%)$ at $t=-0.14$ Myr, which is
half the pulsar age calculated from its timing.

According to Hoogerwerf et al. (2001), the stars HIP 22061 (B2.5V)
and HIP 29678 (B1V) have masses of $8.6M_\odot$ and $11.5M_\odot,$
respectively. They are probably the components of a dissociated
multiple (HIP 29678 is a visual binary) system. We found that out
of the 300 000 orbits, 18 165 encounters occur between these two
stars at separations $\Delta_r\leq10$ pc $(6\%)$; in addition,
5310 of the 18165 encounters occur within $\Delta_r\leq50$ pc of
the cluster $\lambda$Ori at about $t = -1.12$ Myr. As can be seen
from the table, the localization error for each of the objects
under consideration is $\approx20\%.$ This served as a basis for
choosing the size of the neighborhood (50 pc). Therefore, our
results are consistent with the assumption that HIP 22061 and HIP
29678 ran away from the neighborhood of the cluster $\lambda$Ori
about 1.1 Myr ago. Figure 5a shows the expected distribution F3D
of the minimum separation $\Delta_r$ calculated from Eq.~(4) for
the dispersion $\sigma=1.2$ pc. This is an example of very close
encounters; thus, for example, the separation between the two
stars was only $\leq$0.12 pc in two model cases.

The conclusion that HIP 22061 and HIP 29678 are the remnants of a
dissociated binary or multiple system is in agreement with the
analysis of the motion of these stars performed by Hoogerwerf et
al. (2001) using Hipparcos data. However, Hoogerwerf et al. (2001)
questioned the involvement of the star $\lambda$Ori in the event
that led to the breakup of the multiple system HIP 22061--HIP
29678. This conclusion was drawn from the fact that the point on
the celestial sphere with coordinates
$(l,b)=(196^\circ.5,-12^\circ.0)$ was found as a probable current
position of the parent cluster for HIP 22061 and HIP 29678, which
differs by one degree from the current position of the center of
the cluster $\lambda$Ori. The velocity difference effect can be
seen in Fig. 1. More specifically, if the cluster $\lambda$Ori
were located one degree to the left, then all three trajectories
($\lambda$Ori, HIP 22061, and HIP 29678) would intersect at the
same time, $t\approx-1$ Myr.

The kinematic paradox in the motion of the star $\lambda$Ori was
pointed out by Maddalena and Morris (1987). It lies in the fact
that the position of the star $\lambda$Ori in the past does not
coincide with the center of the ring of molecular hydrogen clouds
expanding with a velocity of $14.3\pm2.5$~km s$^{-1}$. We are
talking about a velocity difference of $7-10$ km s$^{-1}$.
According to Maddalena and Morris (1987), the coordinates of the
ring center are $(\alpha,\delta)_{1950}=(5^h 29^m.8\pm1^m.6,
9^\circ 54'\pm 24')$ and, hence, $(l,b)=(194^\circ.7\pm0^\circ.4,
-12^\circ.5\pm0^\circ.4)$ (in our view, these coordinates reflect
the position of the supernova at the time of its explosion). The
position of the ring center remarkably coincides with the point of
intersection of the trajectories for HIP 22061 and HIP 29678 that
we found, namely, $(l,b)=(194^\circ.6\pm0^\circ.2,
-12^\circ.2\pm0^\circ.2).$ The errors here were estimated from the
relation $\varepsilon_{(l,b)} = \arctan(\sigma/r),$ where
$r=438$~pc, and from the dispersion $\sigma=1.2$ pc found above
(Fig.~5a).

Let us assume that the HIP 22061--HIP 29678 pair is a relic of the
system (the cluster $\lambda$Ori) before the time that preceded
the dynamical ``event'' as a result of which the star $\lambda$Ori
received an additional kick $(-10$ km s$^{-1}).$ In this case, we
obtain an independent estimate at the time of the event --- it
could occur no later than 1.12 Myr ago. Moreover, it follows from
the above assumption that the point of intersection of the HIP
22061--HIP 29678 pair indicates the position of the center of the
cluster $\lambda$Ori at a time of $-1.12$ Myr. This suggests that
the supernova explosion that gave rise to the ring of molecular
clouds around the cluster $\lambda$Ori occurred no later than
$-1.12$ Myr.

We can see that the pair of runaway starsHIP22061 and HIP 29678 is
of great interest in studying the evolution of the cluster
$\lambda$Ori. We consider the close encounters of HIP 29678 with
the Geminga pulsar found above as a purely geometric effect. If
both these objects (HIP 29678 and Geminga) resulted from the
evolution of a binary progenitor system, then its trajectory
before its breakup could be arbitrary and the encounter of the HIP
22061--HIP 29678 trajectories would not be as shown in Fig.~1.

\section{DISCUSSION}

Based on an isochronous age estimate for the cluster ASCC21, 12.9
Myr (Kharchenko et al. 2005), we can estimate the minimum mass of
a post-main sequence star. For example, based on Fig.~11.5 from
Sakhibullin (2003) for $\log t = 7.11,$ we obtain an estimate of
the mass for such a star, $\approx15M_\odot.$ On the other hand,
there are no stars bluer than a B1.5V-type star (HIP 25861) in the
cluster ASCC21. Therefore, the more massive B0V-type stars, whose
masses are known to be $\approx15M_\odot,$ can be assumed to have
already left the main sequence. A more detailed comparison with
isochrones, for example, from Schaller et al (1992), also leads to
a similar conclusion.

We may conclude that in the case of ASCC21, the mass of the
probable progenitor star for the Geminga pulsar satisfies well the
condition for the formation of neutron stars with $(10-12)M_\odot
<M<(30-40)M_\odot$ (Zasov and Postnov 2006).

The cluster $\lambda$Ori is considerably younger. Dolan and
Mathieu (2002) estimated the mass and age of the most massive
cluster star, $\lambda$Ori, to be $26.8 M_\odot$ and $\approx5.5$
Myr, respectively. Thus, the higher-mass stars have already
evolved. Therefore, the cluster ASCC21 appears more preferable in
the scenario with a single progenitor star for the Geminga pulsar.

Our simulations showed that there is a probability $(\approx3\%)$
of the simple scenario in which the pulsar was formed in ASCC21 at
$t = -0.52$ Myr and has since moved in the same direction with the
initially received velocity.

As was pointed out by Yakovlev et al. (1999), the characteristic
(i.e., dynamical, determined from timing) age of Geminga was
estimated to within a factor of $\sim$3. Glitches in period, which
are also observed in the Geminga pulsar (Jackson et al. 2002), are
among the factors that affect the dynamical age estimates for
pulsars. On the other hand, the neutron star Geminga exhibits
thermal radiation, which allows its age boundaries to be estimated
by comparison with neutron star cooling models. Thus, for example,
we can see from the data in Table 3 and the cooling curves in Fig.
20 from Yakovlev et al. (1999) that the upper limit for the
Geminga age in the blackbody model reaches $\approx1$ Myr. Similar
estimates of the age boundaries for the Geminga pulsar, $0.1-1$
Myr, can be found in Page et al. (2004).

Thus, the scenario for the formation of Geminga in the cluster
ASCC21 at $t = -0.52$ Myr is consistent with the time intervals
considered. For this scenario to be realized, first, the Geminga
parallax must fall within the range 2--3 mas (Fig. 3) and, second,
there must be traces of the supernova remnant around ASCC21, which
is not yet observed. On the other hand, such a remnant could
disperse completely in $\approx0.5$ Myr.

The assumption that the progenitor star from which the Geminga
pulsar subsequently originated was formed in ASCC21 appears more
realistic. This viewpoint is in agreement with the scenario by
Smith et al. (1994), who believe the Geminga pulsar to be either
the remnant of a runaway OB star ejected from the Orion OB1a
association and exploded as a supernova outside this group or the
result of a supernova explosion in the cluster $\lambda$Ori. Cunha
and Smith (1996) provide arguments that the ring of molecular
hydrogen clouds around the cluster $\lambda$Ori has an age of
300--370 thousand years, in good agreement with the dynamical age
estimate for the neutron star Geminga.

\section*{CONCLUSIONS}

Analysis of the possible encounters of the neutron star Geminga
with open clusters in the Ori OB1a association in the past led us
to conclude that the open cluster ASCC21, a member of the Ori OB1a
association, is the most likely candidate for the birthplace of
either a binary progenitor system that subsequently broke up or a
single progenitor star. Monte Carlo simulations of the encounters
of this pair with Geminga痴 radial velocity $V_r = -100\pm50$ km
s$^{-1}$ showed that close encounters of the pulsar with this open
cluster could occur within $\leq 10$ pc at about $t = -0.52$ Myr.
In addition, the trajectory of the Geminga pulsar passes at a
distance of about 25 pc from the center of the compact OB
association $\lambda$Ori at about $t = -0.39$ Myr.

Previously, the Geminga pulsar was assumed to be related to a wide
neighborhood of the Ori OB1a association, while we found the
specific open cluster (ASCC21) from which the massive progenitor
star or the progenitor binary could be ejected.

In addition, the kinematics of the pair of runaway stars HIP
22061--HIP 29678 under the assumption that they were members of
the cluster $\lambda$Ori in the past allowed us to obtain an
independent estimate for the time of the dynamical ``event'' as a
result of which the star $\lambda$Ori received an additional
kick---it could occur no later than 1.12 Myr ago.

\section*{ACKNOWLEDGMENTS}

We wish to thank the referees for valuable remarks that
contributed to a significant improvement of the paper. This work
was supported by the Russian Foundation for Basic Research
(project no. 08--02--00400) and the ``Origin and Evolution of
Stars and Galaxies'' Program of the Presidium of the Russian
Academy of Sciences and the Program of  State Support for Leading
Scientific Schools of the Russian Federation (grant
NSH--6110.2008.2 ``Multi-wavelength Astrophysical Research'').

 \bigskip
{\bf REFERENCES}
 \bigskip

1. G.F. Bignami and P.A. Caraveo, Nature 357, 287 (1992).

2. G.F. Bignami, P. A.Caraveo, and S. Mereghetti, Nature 361, 704
(1993).

3. A.Blaauw, Bull. Astron. Inst. Netherlands 15, 265 (1961).

4. A.Blaauw, Ann. Rev. Astron. Astrophys. 2, 213 (1964).

5. A.Blaauw,ASP Conf. Ser. 35, 207 (1993).

6. V.V. Bobylev, Pis知a Astron. Zh. 30, 185 (2004) [Astron. Lett.
30, 159 (2004)].

7. C. Brice\~{n}o, N. Calvet, J. Hernandes, et al., Astron. J.
129, 907 (2005).

8. A.G.A. Brown, E.J. de Geus, and P.T. de Zeeuw, Astron.
Astrophys. 289, 101 (1994).

9. D.L. Burch, K.T.S. Brasier, C.E. Fitchel, et al., Nature 357,
357 (1992).

10. P.A. Caraveo, G.F. Bignami, R. Mignani, et al., Astroph. J.
604, 339 (1996).

11. P.A. Caraveo, G.F. Bignami, De Luca, et al., Science 301, 134
(2003).

12. K. Cunha and V.V. Smith, Astron. Astrophys. 309, 892 (1996).

13. W. Dehnen and J.J. Binney, Mon. Not. R. Astron. Soc. 298, 387
(1998).

14. C.J. Dolan and R.D. Mathieu, Astron. J. 118, 387 (1999).

15. C.J. Dolan and R.D. Mathieu, Astron. J. 121, 2124 (2001).

16. C.J. Dolan and R.D. Mathieu, Astron. J. 123, 387 (2002).

17. J. Faherty, F. Walter, and J. Anderson, Astrophys. Space Sci.
308, 225 (2007).

18. P.C. Frisch, Nature 364, 395 (1993).

19. B. Fuchs, D. Breitschwerdt,M.A. Avilez, et al.), Mon. Not. R.
Astron. Soc. 373, 993 (2006).

20. N. Gehrels and W. Chen, Nature 361, 706 (1993).

21. G.A. Gontcharov, Pis知a Astron. Zh. 32, 844 (2006) [Astron.
Lett. 32, 759 (2006)].

22. J.P. Halpern and S.S. Holt, Nature 357, 222 (1992).

23. J.P. Halpern and M. Ruderman, Astrophys. J. 415, 286 (1993).

24. R. Hoogerwerf, J.H.J. de Bruijne, and P.T. de Zeeuw, Astron.
Astrophys. 365, 49 (2001).

25. M.S. Jackson, J.P. Halpern, E. Gotthelf, et al., Astroph. J.
578, 935 (2002).

26. N.V. Kharchenko, A.E. Piskunov, S. S. R\"{o}ser, et al.,
Astron. Astrophys. 440, 403 (2005).

27. N.V. Kharchenko, R.-D. Scholz, A.E. Piskunov, et al., Astron.
Nachr. 328 (2007).

28. P.G. Kulikovsky, Stellar Astronomy (Nauka, Moscow, 1985) [in
Russian].

29. A.D. Kuz知in and V.V. Losovsky, Pis知a Astron. Zh. 23, 323
(1997) [Astron. Lett. 23, 283 (1997)].

30. B. Lindblad, Arkiv Mat., Astron., Fysik, Bd. 20 A, ｹ17 (1927).

31. B. Lindblad, Handbuch der Physik 53, 21 (1959).

32. R.J. Maddalena and M. Morris, Astroph. J. 323, 179 (1987).

33. V.M. Malofeev and O.I. Malov, Nature 389, 697 (1997).

34. P.M. McGehee, Astron. J. 131, 2959 (2006).

35. C. Motch, A.M. Pires, F. Haberl, et al., Astrophys. Space Sci.
308, 217 (2006).

36. D. Page, J. M. Lattimer, M. Prakash, et al., Astroph. J.
Suppl. Ser. 155, 623 (2004).

37. L.J. Pellizza, R.P. Mignani, I.A. Grenier, et al., Astron.
Astrophys. 435, 625 (2005).

38. A.E. Piskunov, E. Schilbach, N.V. Kharchenko, et al., Astron.
Astrophys. 477, 165 (2008).

39. S.B. Popov,M. Colpi, M.E. Prokhorov, et al., Astron.
Astrophys. 406, 111 (2003).

40. N.A. Sakhibullin, Simulation Methods in Astrophysics (Fan,
Kazan, 2003)[in Russian].

41. G. Schaller, D. Schaerer, G. Meynet, et al. Astron. Astroph.
Suppl. Ser. 96, 269 (1992).

42. E. Schilbach and S. S. R\"{o}ser, astro-ph 0806.0762 (2008).

43. Yu. A. Shibanov, S. Zharikov, V. Komarova, et al., Astron.
Astrophys. 448, 313 (2006).

44. V.V. Smith, C. Cunha, and B. Plez, Astron. Astrophys. 281, L41
(1994).

45. van Leeuwen, Astron. Astrophys. 474, 653 (2007).

46. D.G. Yakovlev, K.P. Levenfish, and Yu. A. Shibanov, Usp. Fiz.
Nauk 169, 825 (1999) [Phys. Usp. 42, 737 (1999)].

47. A.V. Zasov and K.A. Postnov, General Astrophysics (Fryazino,
2006) [in Russian].

48. The Hipparcos and Tycho Catalogues, ESA SP-1200 (1997).

\newpage
\begin{figure}[t]
{
\begin{center}
  \includegraphics[width=120mm]{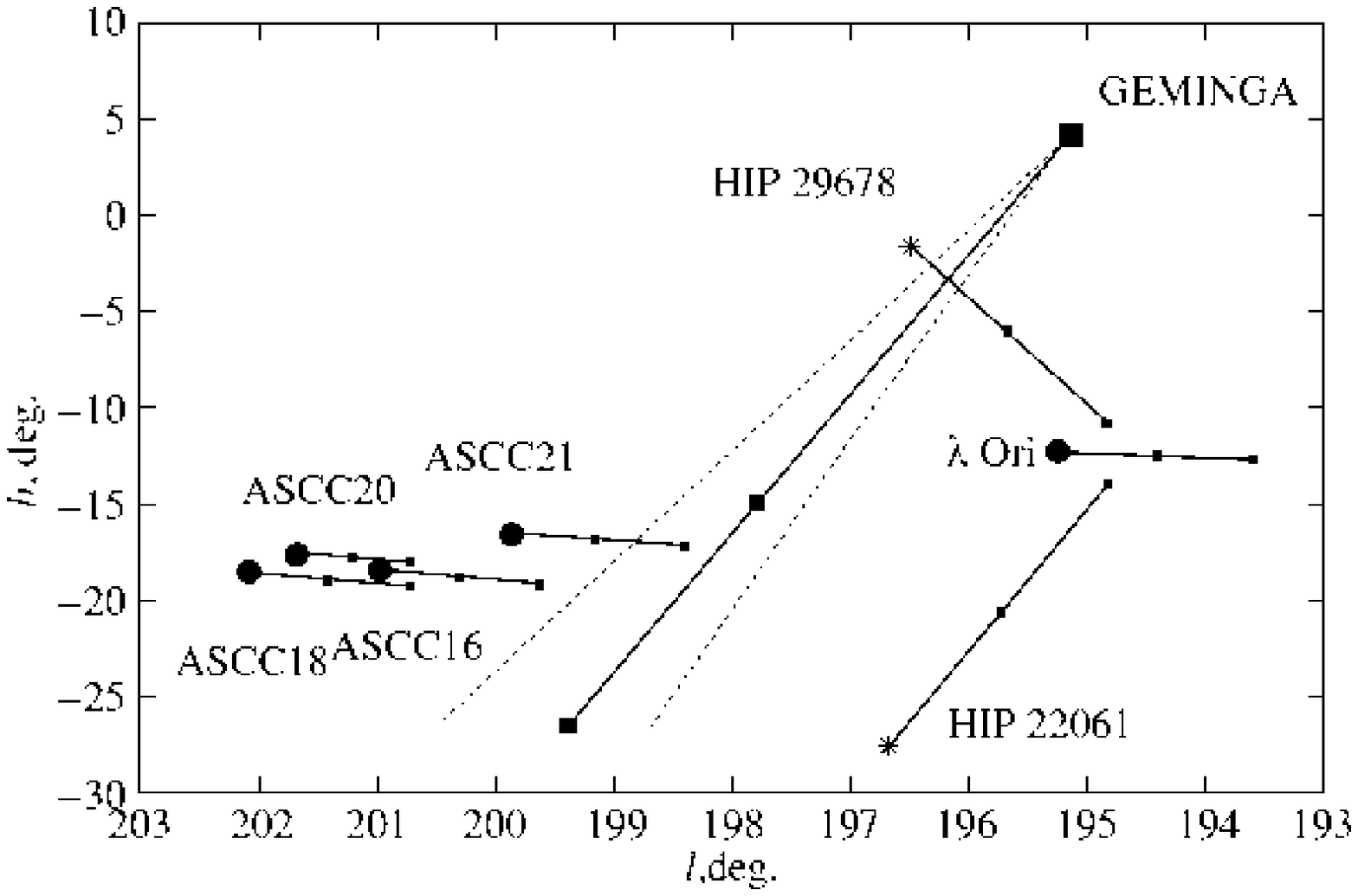}
\end{center}
} Fig.~1. Positions of the open clusters in the Ori OB1
association and the Geminga pulsar (the trajectory was computed
for a radial velocity of $-100$ km s$^{-1}$).
\end{figure}

\begin{figure}[t]
{
\begin{center}
  \includegraphics[width=120mm]{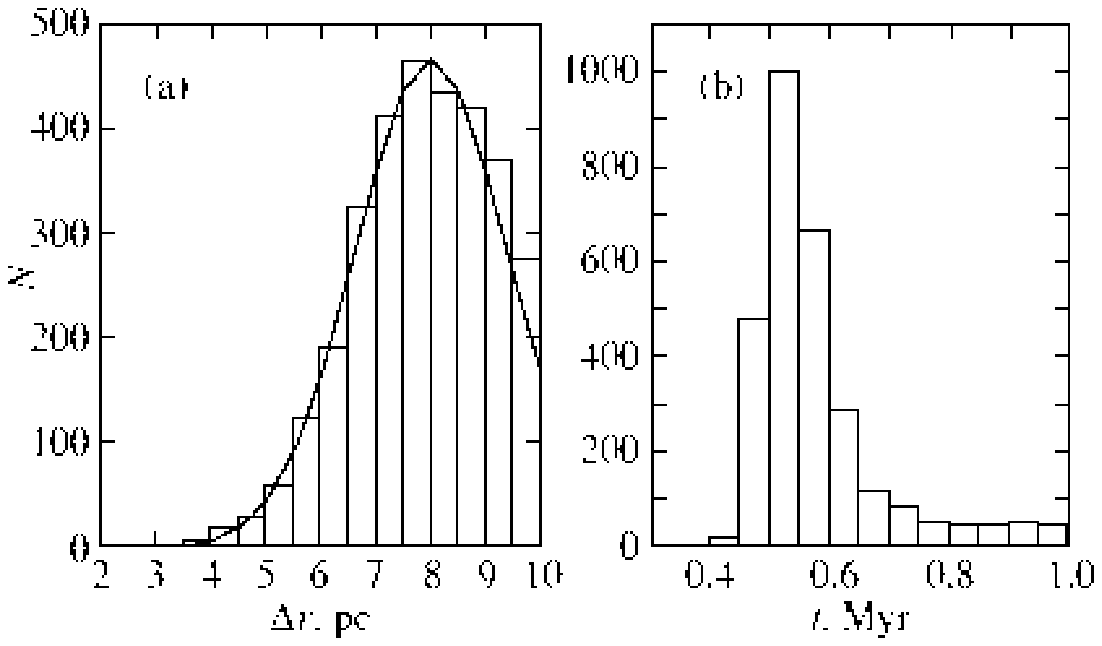}
\end{center}
} Fig.~2. (a) Expected distribution of the minimum separation
$\Delta_r\leq10$ pc for Geminga encounters with the cluster
ASCC21; (b) histogram of encounter times (the time is measured
into the past).
\end{figure}

\begin{figure}[t]
{
\begin{center}
  \includegraphics[width=145mm]{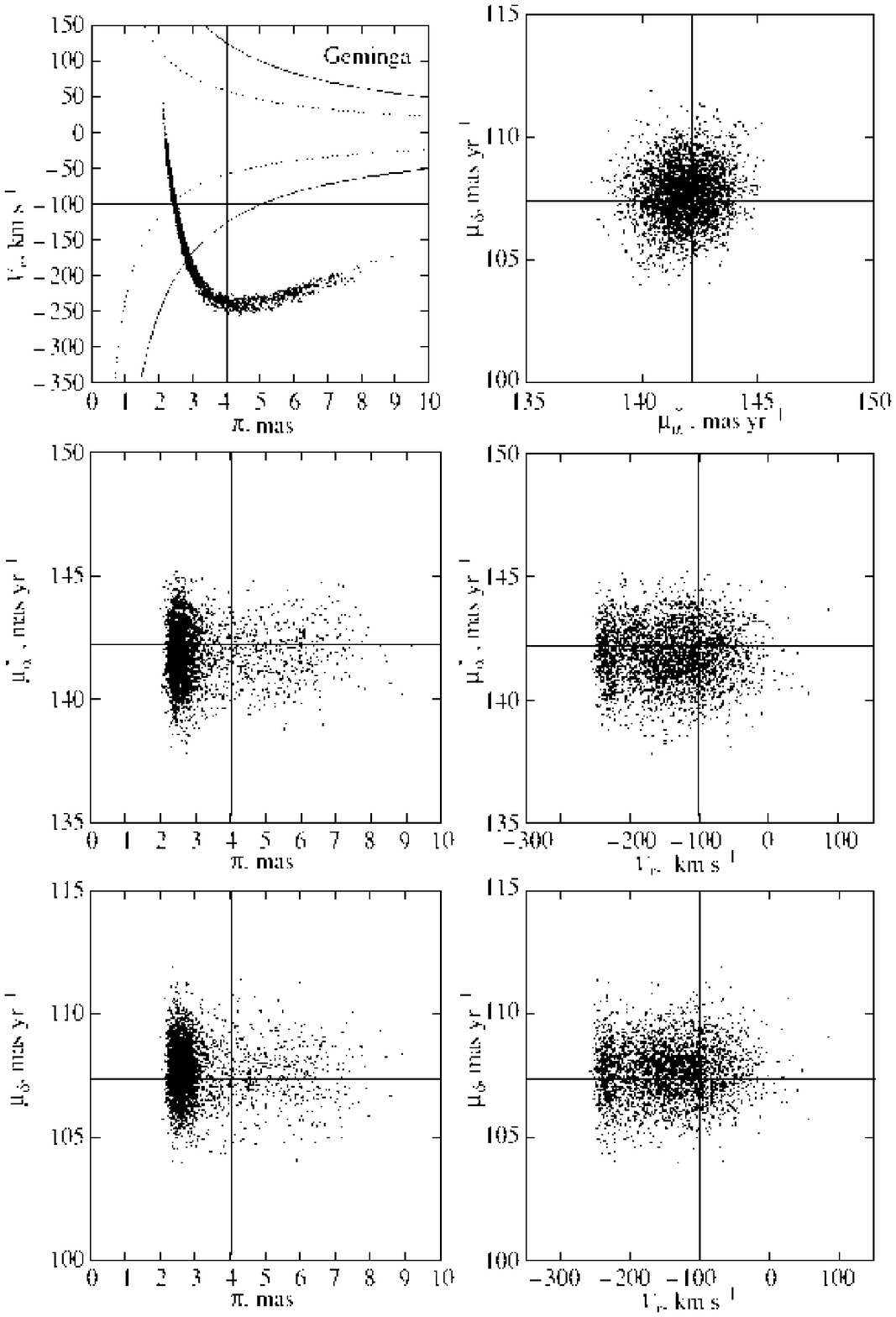}
\end{center}
} Fig.~3. Domains of admissible values for Geminga at which 3137
encounters occur between the pulsar and the cluster ASCC21 to
distances $\Delta_r\leq 10$ pc.
\end{figure}

\begin{figure}[t]
{
\begin{center}
  \includegraphics[width=145mm]{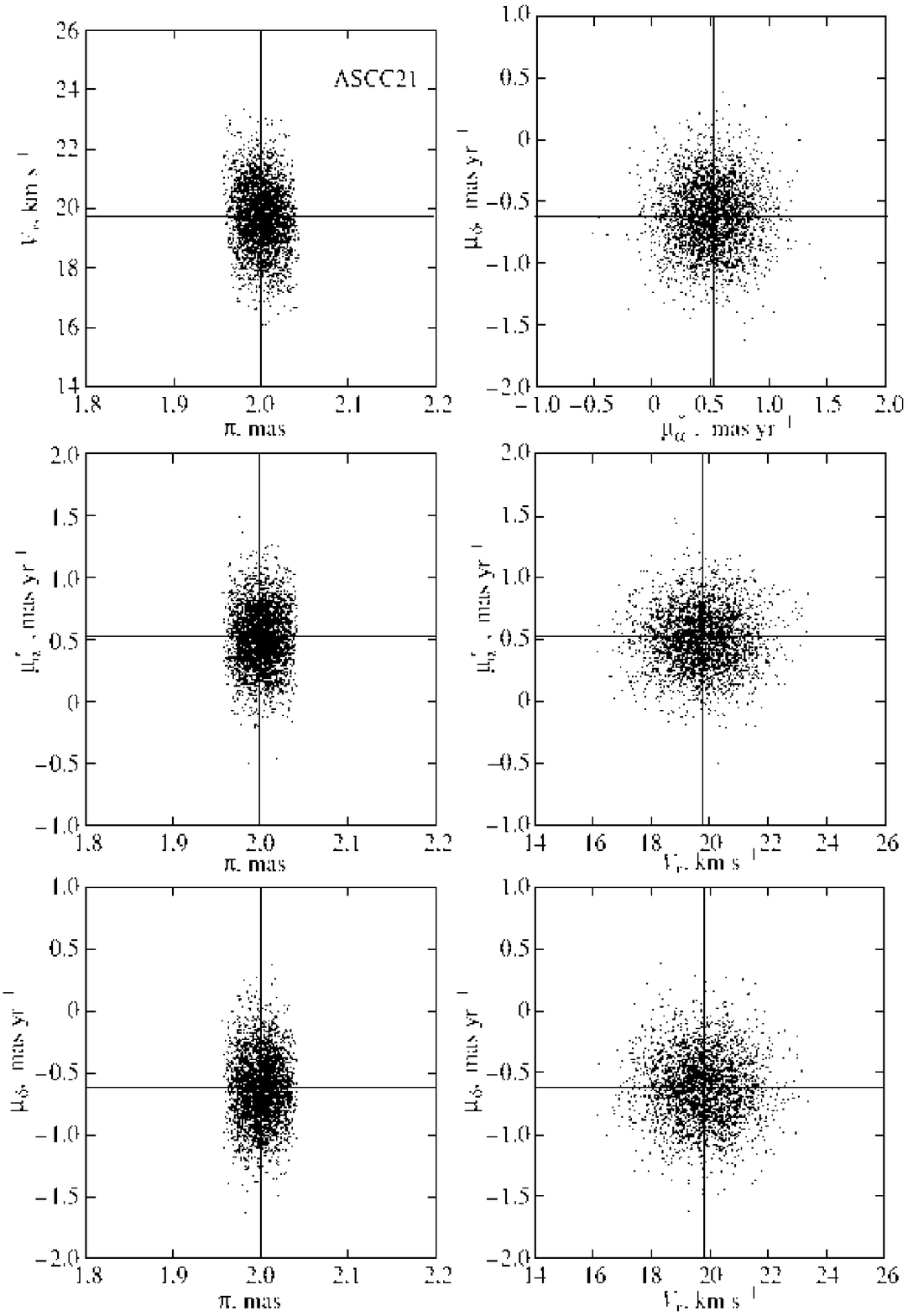}
\end{center}
} Fig.~4.  Domains of admissible values for the cluster ASCC21 at
which 3137 encounters with the pulsar occur at separations
$\Delta_r\leq 10$ pc.
\end{figure}

\begin{figure}[t]
{
\begin{center}
  \includegraphics[width=120mm]{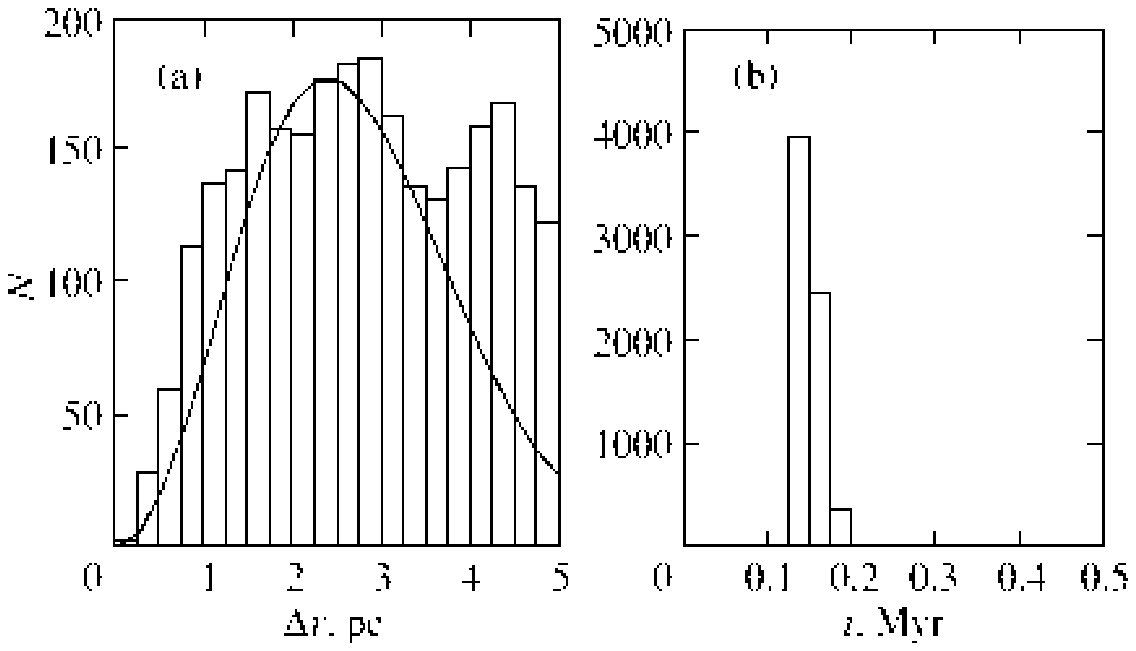}
\end{center}
} Fig.~5.  (a) Expected distribution of the minimum separation
$\Delta_r\leq 10$ pc for the encounters of HIP 22061 and HIP 29678
within $\Delta_r\leq 50$ pc of the cluster $\lambda$Ori; (b)
histogram of encounter times (the time is measured into the past).
\end{figure}

\end{document}